\documentclass{PoS}

\usepackage{epsfig,multicol}

\newcommand{\rc}{\renewcommand}
\newcommand{\nc}{\newcommand}
\nc{\ba}{\begin{eqnarray}}
\nc{\ea}{\end{eqnarray}}
\rc{\cal}{\mathcal}
\nc{\eq}[1]{Eq.~(\ref{#1})}
\nc{\eqs}[2]{Eqs.~(\ref{#1}--\ref{#2})}
\nc{\eqss}[2]{Eqs.~(\ref{#1},\ref{#2})}
\nc{\se}[1]{Sec.~\ref{#1}}
\nc{\re}[1]{Ref.~\cite{#1}}
\nc{\fig}[1]{Fig.~\ref{#1}}
\nc{\rmi}[1]{{\mbox{\scriptsize #1}}}
\nc{\e}{\epsilon}
\rc{\a}{\alpha}
\nc{\ta}{{\tilde\alpha}}
\nc{\tb}{{\tilde\beta}}
\nc{\fr}{\frac}
\def\({\left(}
\def\){\right)}
\nc{\lk}{\left[}
\nc{\rk}{\right]}
\nc{\lb}{\left\{}
\nc{\rb}{\right\}}
\nc{\ld}{\left.}
\nc{\rd}{\right.}
\nc{\hP}{{\hat\Pi}}
\nc{\hg}{{\hat g}}
\nc{\hhg}{{\tilde g}}
\nc{\hm}{{\hat m}}
\nc{\hb}{{\hat\beta}}
\nc{\hl}{{\hat\lambda}}
\nc{\hn}{{\hat n}}
\nc{\nn}{\nonumber\\}
\nc{\order}{{\cal O}}
\nc{\sumint}{\sum\!\!\!\!\!\!\int}
\nc{\msbar}{$\overline{\rm MS}$}
\nc{\lmsbar}{{\Lambda_{\overline{\rm MS}}}}
\nc{\lms}{\Lambda_{\overline{\mbox{\tiny MS}}}}
\nc{\tr}{\mbox{Tr}}
\nc{\pSB}{{p_{\rmi{SB}}}}
\nc{\Nc}{{N_{\rmi{c}}}}
\nc{\Nf}{{N_{\rmi{f}}}}

\title{Weak-coupling expansion of the hot QCD pressure}

\ShortTitle{Weak-coupling expansion of the hot QCD pressure}

\author{\speaker{York~Schr\"oder}\\
	Faculty of Physics, University of Bielefeld,
        D-33501 Bielefeld, Germany\\
	E-mail: \email{yorks@physik.uni-bielefeld.de}}

\abstract{%
We review recent progress made in determining the pressure of hot QCD up
to the first non-perturbative term in its weak-coupling expansion.\\[14mm]
\mbox{}\hfill BI-TP 2006/08, hep-ph/0605057}

\FullConference{29th Johns Hopkins Workshop on current problems in 
particle theory: strong matter in the heavens\\1-3 August\\Budapest, Hungary}

\begin{document} 


\section{Introduction}

The pressure of hot QCD, $p_{\rm QCD}$, 
constituting one of the most fundamental
thermodynamic observables, has been under theoretical study for 
several decades now. 
Being of fundamental importance to cosmology (due to its influence
on the cooling rate of the early universe) as well as of potential
relevance to heavy ion collisions (through its influence on the 
thermodynamic expansion rate),
$p_{\rm QCD}$ has been computed using a variety of methods,
including lattice Monte-Carlo, weak-coupling and 
large-$\Nf$-methods, to name a few.

Although at large temperatures $T$ asymptotic freedom leads to the 
expectation that weak-coupling methods are sufficient to accurately 
describe the deconfined phase, it is well known that the infrared (IR)
sector of QCD produces a challenge for perturbation theory,
which for the case of the pressure arises at order $g^6T^4$
\cite{Linde}.

It has however been realized that this challenge can be overcome 
in an effective theory setup, where the problematic sector is
described by a dimensionally reduced theory 
\cite{Appelquist:1981vg}. The key observation is that thermal
(equilibrium) QCD possesses three distinct physical scales, two
of them generated dynamically. The contributions to the pressure
(and to any other thermodynamic observable) from each of these scales 
can be obtained from carefully constructing and matching a series of 
effective theories 
(for details, see e.g. \cite{Braaten:1995ju,Kajantie:2002wa}). 

The theories under consideration are (hard) QCD, electrostatic QCD
(EQCD) and magnetostatic QCD (MQCD), governing physics on length scales 
$1/T$ (the typical scale for a particle in the heat-bath), 
$1/gT$ (the dynamically generated screening length for longitudinal 
gluonic excitations) and 
$1/g^2T$ (the dynamically generated screening length for transverse 
gluonic excitations), respectively.
While the first two are amenable to perturbative calculations,
MQCD is purely non-perturbative and has to be treated on the lattice.
Viewing the gauge coupling $g(T)$ as parametrically small (which is
certainly justified at asymptotically high temperatures), these
three scales are well separated, and can hence be dealt with individually
via the effective theory setup. Schematically, for the pressure one can
write $p_{\rm QCD}=p_E+p_M+p_G$, where each contribution depends
on the matching scales. This scale-dependence will cancel in the sum,
rendering $p_{\rm QCD}$ a well-defined physical observable.

In this letter, we will give
a somewhat condensed account of what we currently know about 
the different contributions to $p_{\rm QCD}$.


\section{Status of the QCD pressure}

Below, we specify the contributions to the \msbar~pressure 
$p_{\rm QCD}=p_G+p_M+p_E$ \cite{Braaten:1995ju} from
each physical scale individually, for the case of 
gauge group SU($\Nc$) and $\Nf$ quark flavors.
We will work at zero quark masses $m_{q_i}=0$ and vanishing 
chemical potentials $\mu_f=0$,
and display all dependence on the \msbar~scale 
$\bar\mu^2 = 4\pi e^{-\gamma_0}\mu^2$ by 
$L\equiv\ln\fr{\bar\mu}{4\pi T}$.
Effects due to finite quark masses \cite{Kapusta:1979fh,Laine:2006cp} 
and chemical potentials \cite{chemPot} 
are available in the literature, but will not be discussed here.


\nc{\pert}{[\mbox{\bf pert}]}
\nc{\nspt}{[\mbox{\bf nspt}]}
\nc{\nonpert}{[\mbox{\bf non-pert}]}

\subsection{Contributions from the ultra-soft scale $g^2 T$, i.e. from MQCD}

Ultra-soft physics is not accessible by perturbative methods, due to
the unscreened transverse gluonic sector, which would lead to severe
infrared problems \cite{Linde}.
This sector is governed by a three-dimensional pure gauge theory.
Its only parameter is the dimensionful 3d gauge coupling $g_M^2$, which 
we write as $\hg_M^2\equiv\frac{\Nc g_M^2}{16\pi^2 T}$.
The screening length gets generated non-perturbatively, making a numerical
lattice Monte-Carlo treatment necessary. 
The detailed setup for how to incorporate the ultra-soft contribution
into the physical pressure by a carefully defined mixture of 
perturbative and non-perturbative coefficients 
is explained in detail in \re{Laine:2006cp}. The result is
\ba \label{eq:pG}
\fr{p_G(T)}{\mu^{-2\e}} &=&
 d_A 16\pi^2 T^4 \hg_M^6 \lk 8\a_G\(\fr1{8\e}+L+\ln\frac{\Nc^2}{4\pi\hg_M^2}+1\) 
 +\frac13
\rd\nonumber\\&+&\ld
\pert
-\nspt
+\nonpert
 +\order(\e) \vphantom{\frac12}\rk , 
\ea
where $d_A=\Nc^2-1$, and
$\a_G=\fr{43}{96}-\fr{157}{6144}\pi^2$ \cite{Kajantie:2002wa,Schroder:2003uw}
is a perturbative 4-loop coefficient.
The three coefficients enclosed in square brackets originate from 
measuring the 3d YM pressure
on the lattice and matching the result to the \msbar~scheme.
To be more precise, they are the following.

$\bullet$ The first number stems from a non-perturbative lattice Monte-Carlo 
measurement of the 3d plaquette in pure SU($\Nc$) theory 
\cite{Hietanen:2004ew}, 
\ba \label{eq:nonpert}
\nonpert &=& \frac{(4\pi)^4}{8d_A\Nc^6} \lim_{\beta\rightarrow\infty}\lb
\beta^4\left\langle 1-\frac1{\Nc} \tr P \right\rangle_a 
-\lk c_1\beta^3+c_2\beta^2+c_3\beta+c_4\ln\beta\rk
\rb\\ \label{eq:nonpertRes}
&=& 10.7(4) \quad \mbox{at $\Nc=3$} \;,\nonumber
\ea
where $\beta=\frac{2\Nc}{g_M^2 a}$ denotes the dimensionless lattice coupling,
and $c_{1..4}$ are divergences of the 3d lattice-regularized plaquette
which can be computed in lattice perturbation theory.
They read
\ba
c_1 &=& \frac{d_A}{3}\;,\\
c_2 &=& \frac{d_A}{(4\pi)^2}\(-\frac{8\pi^2}9 +5.25449 \Nc^2\)\;,\\
c_3 &=& d_A\([0.04978944(1)]+[-0.04289464(7)]\Nc^2
+[0.0147397(3)]\Nc^4\) \\
&=& 6.8612(2) \quad\mbox{at $\Nc=3$}\;,\nonumber\\
c_4 &=& \frac{d_A\Nc^6}{(4\pi)^4}\,64\a_G\;.
\ea
The number in $c_2$ is a sum of typical 2-loop (infinite--volume)
lattice integrals \cite{Heller:1984hx,Hietanen:2004ew},
\ba
-\frac23\(\frac{\Sigma^2}4-\pi\Sigma
-\frac{\pi^2}2+4\kappa_1+\frac23\kappa_5\) &\approx& 5.25449 \;,
\ea
with ($K$ is the complete elliptic integral of first kind) 
\cite{Farakos:1994xh,Laine:1997dy,Hietanen:2004ew}
\ba
\Sigma &=& \frac1{\pi^2}\int_{0}^{\pi}{\rm d}^3x 
\frac1{\sum_i \sin^2x_i} \\
&=& 
\frac8{\pi}\(18+12\sqrt2-10\sqrt3-7\sqrt6\)
K^2\((2-\sqrt3)^2(\sqrt3-\sqrt2)^2\) 
\;\approx\; 3.1759114 \;,\\
\kappa_1 &=& \frac1{4\pi^4}\int_{-\pi/2}^{\pi/2}{\rm d}^3x{\rm d}^3y
\frac{\sum_i \sin^2x_i \sin^2(x_i+y_i)}
{\sum_i\sin^2x_i \sum_i\sin^2(x_i+y_i) \sum_i\sin^2y_i} 
\;\approx\; 0.958382(1) \;,\\
\kappa_5 &=& \frac1{\pi^4}\int_{-\pi/2}^{\pi/2}{\rm d}^3x{\rm d}^3y
\frac{\sum_i \sin^2x_i \sin^2(x_i+y_i) \sin^2(y_i)}
{\sum_i\sin^2x_i \sum_i\sin^2(x_i+y_i) \sum_i\sin^2y_i}
\;\approx\; 1.013041(1) \;.
\ea
The coefficient $c_3$ has been estimated by numerical stochastic
perturbation theory (NSPT) for $\Nc=3$ 
\cite{DiRenzo:2004ws} and, with higher numerical accuracy and full $\Nc$
dependence, computed from 3-loop diagrams 
in lattice perturbation theory \cite{Panagopoulos:2006ky}. 
In principle, it would be nice to know the full $\Nc$\/-dependence 
of \eq{eq:nonpert}.

$\bullet$ The second number stems from an estimation of (the sum of all)
4-loop vacuum diagrams in lattice perturbation theory by NSPT
\cite{DiRenzo:2004ge}, with the 
IR divergence regulated by massive gluon- and ghost-propagators
(mass term $\frac{m^2}2 A^2$ and $m^2 \bar cc$ in the action)
\cite{Torrero:2005na}.
NSPT works on a finite lattice of volume $(aL)^3$, so the infinite--volume
limit has to be taken first to ensure that the IR is regulated 
by the mass term only. 
The result is \cite{TorreroThesis,NEW}
\ba
\nspt &=& \frac{(4\pi)^4}{8d_A\Nc^6}\,
\lim_{am\rightarrow0}\lim_{L\rightarrow\infty}
\lb\left.\left\langle 1-\frac1{\Nc} \tr P
\right\rangle_{am}\right|_{\mbox{$\beta^{-4}$ term}}
-c_4\ln\frac1{am}\rb \\
&=& \frac{(4\pi)^4}{8d_A\Nc^6} \lk c_{40}' +c_{41}' \Nc^2 
+c_{42}' \Nc^4 +c_{43}' \Nc^6 \rk \nonumber\\ \label{eq:nsptRes}
&\approx& \frac{4\pi^4}{729}\, 25.8(8) \;=\; 13.8(4) \quad \mbox{in Feynman gauge at $\Nc=3$} 
\;.\nonumber
\ea
Note that the errorbar of this number matches
the precision obtained for $\nonpert$. 
It would be nice to know all four coefficients,
in Feynman gauge, 
either by a direct diagrammatic evaluation,
or by doing NSPT for (at least) four different values of $\Nc$.

$\bullet$ The third number stems from a matching 4-loop computation
in the $(3-2\e)$d continuum theory, regulated in the IR by gluon- 
and ghost-masses, with gauge parameter $\xi$.
Gauge dependence, introduced by the IR regulator, is guaranteed to
cancel against that in $\nspt$. The result reads
\ba
\sum\lk\mbox{4loop YM vac diags}\rk &=& g^6d_A\Nc^3\(\frac1m J\)^4
\lb\frac{\a_G}{\e}+\pert+\order(\e)\rb
\ea
where $J$ is the 1-loop massive tadpole integral $\int_p\, 1/(p^2+m^2)$.
We choose Feynman gauge $\xi=1$
which here leads to modified propagators $1/p^2\rightarrow 1/(p^2+m^2)$,
and obtain \cite{YS:unpublished} 
\ba
\pert = -3.73134481146281478501 \quad \mbox{in Feynman gauge}
\ea
where the number can be expressed 
in terms of 18 fully massive
4-loop scalar master integrals \cite{Schroder:2002re,Schroder:2003kb}. 
For general $\xi$, we would have to calculate vacuum diagrams with
two mass scales ($m^2$ and $\xi m^2$), which presently is beyond 
our computational capabilities.

The matching condition for the 3d gauge coupling reads
\cite{Farakos:1994kx,Giovannangeli:2003ti,Laine:2005ai}
\ba
\label{eq:gM}
 \hg_M^2\equiv\fr{\Nc g_M^2}{16\pi^2 T} = 
 \hg_E^2 \lk 1 -\frac1{12}\,\frac{\hg_E^2}{\hm_E} 
 -\frac{17}{288}\,\frac{\hg_E^4}{\hm_E^2} 
 -\frac{2\!-\!\hn}{24}\,\frac{\hg_E^2\hl_E^{(1)}}{\hm_E^2} 
 -\frac{3\hn\!-\!1}{24}\,\frac{\hg_E^2\hl_E^{(2)}}{\hm_E^2} 
 \rk 
 +\!\order\!\(\!\frac{\hg_E^8}{\hm_E^{3}}\!\) \,,
\ea
where $\hn\equiv\fr{\Nc^2-1}{\Nc^2}$. 
For the $g^6$ pressure, only the leading coefficient is relevant.


\subsection{Contributions from the soft scale $g T$, i.e. from EQCD}

Soft-scale physics is governed by a three-dimensional gauge theory,
coupled to an adjoint Higgs field. This adjoint Higgs theory possesses
a small number of dimensionful coupling constants, which are related
to the parameters of full QCD (being $g^2$ and $T$) by the equations 
given below.
The contribution of this sector to the pressure is given by
\ba \label{eq:pM}
\fr{p_M(T)}{\mu^{-2\e}} &=&
 d_A 16\pi^2 T^4 \lb 
 \hm_E^3\lk \fr13 +\order(\e) \rk
\rd\nn&+&{}\ld
 \hg_E^2\hm_E^2\lk -\fr1{4\e} +\(-L+\fr12\ln\hm_E^2+\ln2-\fr34\) 
 +\order(\e) \rk
\rd\nn&+&{}\ld
 \hg_E^4\hm_E\lk \(-\fr{89}{24}-\fr{\pi^2}6+\fr{11}6\ln2\) +\order(\e) \rk 
\rd\nn&+&{}\ld
 \hg_E^6\lk \a_M\(\fr1\e+8L-4\ln\hm_E^2-8\ln2\) +\beta_M +\order(\e) \rk 
\rd\nn&+&{}\ld
 \hl_E^{(1)}\hm_E^2\lk \fr{\hn-2}4 +\order(\e) \rk 
 +\hl_E^{(2)}\hm_E^2\lk \fr{1-3\hn}4 +\order(\e) \rk 
\rd\nn&+&{}\ld
 \order(\hg_E^8 \hm_E^{-1},\hl_E^2\hm_E) 
\rb ,
\ea
with the 4-loop coefficients 
$\a_M = \fr{43}{32}-\fr{491}{6144}\pi^2$,
$\beta_M =
 -\frac{311}{256}-\frac{43}{32}\ln2-\frac{19}6\ln^22+\frac{77}{9216}\pi^2
 -\frac{491}{1536}\pi^2\ln2+\frac{1793}{512}\zeta(3)+\gamma_{10}=
 -1.391512$ \cite{Kajantie:2003ax}, 
where $\gamma_{10}$ is the leading coefficient of a finite 
3d scalar 4-loop integral that is known numerically only 
\cite{Kajantie:2003ax},
\ba
\gamma_{10} &=& (4\pi)^4 
\int_{-\infty}^\infty\frac{{\rm d}^3x_1}{(2\pi)^3}
\frac{{\rm d}^3x_2}{(2\pi)^3}
\frac{{\rm d}^3x_3}{(2\pi)^3}
\frac{{\rm d}^3x_4}{(2\pi)^3}\,  
\frac1{(x_1-x_3)^2}\, \frac1{(x_2-x_3)^2}
\times \nonumber\\&\times&
\frac1{x_1^2+1}\, \frac1{x_2^2+1}\, \frac1{(x_1-x_4)^2+1}\, 
\frac1{(x_2-x_4)^2+1}\, \frac1{(x_3-x_4)^2+1}\\
&=& 0.171007009753(1) \;,
\ea
and the matching parameters are \cite{Farakos:1994xh,Braaten:1995ju}
\ba
 \label{eq:mE}
 \hm_E^2 \equiv 
 \(\fr{m_E}{4\pi T}\)^2 &=& 
 \hg^2 \lk \ta_{\rm E4} +\(2\ta_{\rm E4}L+\ta_{\rm E5}\)\e +\order(\e^2) \rk 
\\ &+&
 \hg^4\lk \( 2\hb_0\ta_{\rm E4} L+\ta_{\rm E6} \) 
 +(6\hb_0\ta_{\rm E4}L^2+\tb_{\rm E2}^{(L)}L+\tb_{\rm E2})\e 
 +\order(\e^2) \rk
 +\order(\hg^6) \,, \nonumber\\
\label{eq:gE}
\hg_E^2 \equiv
 \fr{\Nc g_E^2}{16\pi^2 T} &=& 
 \hg^2 
 + \hg^4\lk \( 2\hb_0 L +\ta_{\rm E7} \) 
 +(2\hb_0 L^2+2\ta_{\rm E7}L+\tb_{\rm E3})\e 
 +\order(\e^2) \rk \nn&+&{}
 \hg^6\lk 4\hb_0^2L^2 +2\(\hb_1+2\hb_0\ta_{\rm E7}\)L 
 +\tilde\gamma_{\rm E1}+\order(\e)\rk 
 +\order(\hg^8) \,, \\
\hl_E^{(1)} \equiv 
 \fr{\Nc^2 \lambda_E^{(1)}}{16\pi^2 T} &=& 
 \hg^4\lk 4 +\order(\e) \rk +\order(\hg^6) \;, \\
\hl_E^{(2)} \equiv
 \fr{\Nc \lambda_E^{(2)}}{16\pi^2 T} &=& 
 \hg^4\lk \fr43(1-z) +\order(\e) \rk +\order(\hg^6) \;,
\ea
where we have used the beta-function coefficients 
$\hb_0=\fr{11-2z}3$,
$\hb_1=\fr{34}3-\fr{10}3z-z\hn$ and,
for brevity, set $z\equiv \Nf/\Nc$.
Writing 
$Z_n\equiv\zeta^\prime(-n)/\zeta(-n)$, 
the coefficients read \cite{Braaten:1995ju,???,Huang:1994cu}
\ba
\label{eq:aE4}
\ta_{\rm E4} &=& \fr{2+z}6 \;,\\ 
\ta_{\rm E5} &=& 2\ta_{\rm E4}Z_1 +\fr{z}6(1-2\ln2) \;,\\
\label{eq:aE6}
\ta_{\rm E6} &=& \fr13\ta_{\rm E4}(6\hb_0\gamma_0+5+2z-8z\ln2) 
 -\fr{z}2\hn \;,\\
\ta_{\rm E7} &=& 2\hb_0\gamma_0 +\fr13 -\fr83z\ln2 \;,
\ea
as well as \cite{Schroder:2004gt,Laine:2005ai}
\ba
\tb_{\rm E2}^{(L)} &=& 
 4\hb_0\ta_{\rm E4}(2\gamma_0+Z_1)
 +\fr19(20+29z+2z^2)
-
 2z\(\hn+3\ln2\)
 -\fr43 z^2 \ln2 \;, \\
\tb_{\rm E2} &=& 
 \fr14\hb_0\ta_{\rm E4}\(+\pi^2-16\gamma_1\)
 +\fr23\ta_{\rm E4}Z_1(6\hb_0\gamma_0+5+2z-8z\ln2)
 \nn&+&{}
 \fr29\gamma_0(5+10z-(19+2z)z\ln2)
 +\fr29
 +\fr{z}{18}(7+6\ln2-16\ln^22)
 \nn&+&{}
 \fr{z^2}9(1-2\ln2+4\ln^22)
 -\fr{z}6\hn(3+6\gamma_0+6Z_1+10\ln2) \;,\\
\tb_{\rm E3} &=&
 \(\frac{\pi^2}4-4\gamma_1\)\hb_0+\frac23\gamma_0
 -\frac83\ln2(\ln2+2\gamma_0)z \;,\\
\tilde\gamma_{\rm E1} &=& 2\hb_1 \gamma_0 +\ta_{\rm E7}^2
 +\frac{341}{18}-\frac{10}{9}\zeta(3)
\nonumber\\&-&{}
 \frac{z}9(43+24\ln2+5\zeta(3))
 -\frac{z}{12}\hn(23+80\ln2-14\zeta(3)) \;,
\ea
where the $\gamma_n$ are expansion coefficients of the Zeta function
$\zeta(1-\e)=-\frac1\e+\sum_{n=0}^\infty \frac{\e^n}{n!}\,\gamma_n$
(note that $\gamma_0\equiv\gamma_E=0.577216$).
For the $g^6$ pressure, the $\hg^6$ terms of \eq{eq:gE}
are irrelevant.


\subsection{Contributions from the hard scale $2\pi T$, i.e. from thermal QCD}

Hard-scale physics can be treated perturbatively,
in a simple $g^2$\/-expansion, without the need for resummations,
thermal masses, or hard thermal loops. This is due to all IR effects 
being properly incorporated into EQCD and MQCD, and is one of the main 
conceptual advantages of using the effective theory setup.
The contribution to the pressure from hard momentum scales reads
\ba \label{eq:pE}
\fr{p_E(T)}{\mu^{-2\e}} &=&
 d_A 16\pi^2 T^4 \fr1{16}\fr1{45} \lb \vphantom{\frac12}
 \ta_{\rm E1} 
 +\hg^2\lk \ta_{\rm E2} +\order(\e) \rk 
\rd\nn&+&{}\ld
 \hg^4\lk \ta_{\rm E4}\fr{180}{\e} 
 +(180\cdot 6\ta_{\rm E4}+2\hb_0\ta_{\rm E2} )L +\ta_{\rm E3} +\order(\e) \rk 
\rd\nn&+&{}\ld
 \hg^6\lk 
 \fr{\tb_{\rm E1}^{\rm (div)}}{\e} 
 +\tb_{\rm E1}^{(L^2)}L^2
 +\tb_{\rm E1}^{(L)}L
 +\tb_{\rm E1} 
 +\order(\e) \rk
 +\order(\hg^8) \rb ,\quad
\ea
with ideal-gas coefficient $\ta_{\rm E1} = 1 +\fr74\fr{z}{\hn}$, 
$\ta_{\rm E2} = -\fr54(4+5z)$ \cite{Shuryak:1977ut,Chin:1978gj}, 
and \cite{Arnold:1994ps}
\ba
\ta_{\rm E3} &=& 
 180(\ta_{\rm E4})^2\gamma_0
 +5\lk \(\fr{116}5+\fr{220}3Z_1-\fr{38}3Z_3\) 
\rd\nn&+&{}\ld
 \fr{z}2\(\fr{1121}{60}-\fr{157}5\ln2+\fr{146}3Z_1-\fr13Z_3\) 
\rd\nn&+&{}\ld
 \fr{z^2}4\(\fr13-\fr{88}5\ln2+\fr{16}3Z_1-\fr83Z_3\) 
 +\fr{z}4\hn\(\fr{105}4-24\ln2\) \rk ,\quad
\ea
and unknown coefficients $\beta_{\rm E1}$, which can be determined
e.g. by a 4-loop computation of vacuum diagrams in thermal QCD.
Since $p_{\rm QCD}$ is physical, 
the divergent and scale-dependent parts of $\beta_{\rm E1}$
are related to the other coefficients introduced in the above,
serving as a valuable check on this open computation.
Specifically, from 2-loop running of the 4d gauge coupling
\ba
\hg^2 \equiv \fr{\Nc g^2(\bar\mu)}{16\pi^2} &=&
 \hg^2(\bar\mu_0) +\hg^4(\bar\mu_0)(-2\hb_0\ell) 
 +\hg^6(\bar\mu_0)(4\hb_0^2\ell^2-2\hb_1\ell) \;,
\ea
where 
$\ell \equiv \ln\fr{\bar\mu}{\bar\mu_0} = L-\ln\fr{\bar\mu_0}{4\pi T}$,
one can already fix
\ba
\tb_{\rm E1}^{\rm (div)} &=& 
 180\lk 4\hb_0\ta_{\rm E4} L +\ta_{\rm E6} +\ta_{\rm E4}\ta_{\rm E7}
 -4(\a_G+\a_M) \rk , \\
\tb_{\rm E1}^{(L^2)} &=& 
 180\lk 28\hb_0\ta_{\rm E4} \rk +4\hb_0^2\ta_{\rm E2} \;, \\
\tb_{\rm E1}^{(L)} &=& 
 180\lk 4\ta_{\rm E6} +8\ta_{\rm E4}\ta_{\rm E7} -2\hb_0\ta_{\rm E5} 
 -32(\a_G+\a_M) +\tb_{\rm E2}^{(L)} \rk 
\nn&+&{}
 2\hb_1\ta_{\rm E2} +4\hb_0\ta_{\rm E3} \;. 
\ea
The remaining $g^6$\/-coefficient, $\tb_{\rm E1}$ however, entails
a four-loop computation of all connected vacuum diagrams involving
quarks, gluons and ghosts, a computation that has so far not been
tackled due to the formidable task of solving many genuine 4-loop
sum-integrals. From diagrammatic arguments, it is clearly
a polynomial in $z=\Nf/\Nc$,
\ba
\tb_{\rm E1}=\#_0+z\#_1+z^2\#_2+z^3\#_3
\;,
\ea
and we will in the following indicate how two of its coefficients 
(the first and last) can be crudely estimated numerically already.


\section{Putting everything together} \label{se:choices}

Expanding in $\e$, all poles cancel, as they should. 
In practice we make use of Eqs.~(\ref{eq:gM}),(\ref{eq:mE}) and (\ref{eq:gE}) 
to re-expand all terms with a factor $1/\e$ or $L$ 
in Eqs.~(\ref{eq:pG}) and (\ref{eq:pM}) in terms of $\hg^2$.
After cancellation of the poles (and taking into account 
terms like $\frac1\e\cdot\e$), we can now take the limit 
$\e\rightarrow0$ in Eqs.~(\ref{eq:mE}), (\ref{eq:gE}), whence
\ba
 \hm_E^2 &=& 
 \hg^2 \ta_{\rm E4} 
 +\hg^4\lk 2\hb_0\ta_{\rm E4} L+\ta_{\rm E6} \rk
 +\order(\hg^6) \,, \\
\hg_E^2 &=& 
 \hg^2 
 + \hg^4\lk 2\hb_0 L +\ta_{\rm E7} \rk
 +\hg^6\lk 4\hb_0^2L^2 +2\(\hb_1+2\hb_0\ta_{\rm E7}\)L 
 +\tilde\gamma_{\rm E1} \rk 
 +\order(\hg^8) \,. 
\ea 

\begin{figure}
\centerline{%
\epsfxsize=7.4cm\epsfbox{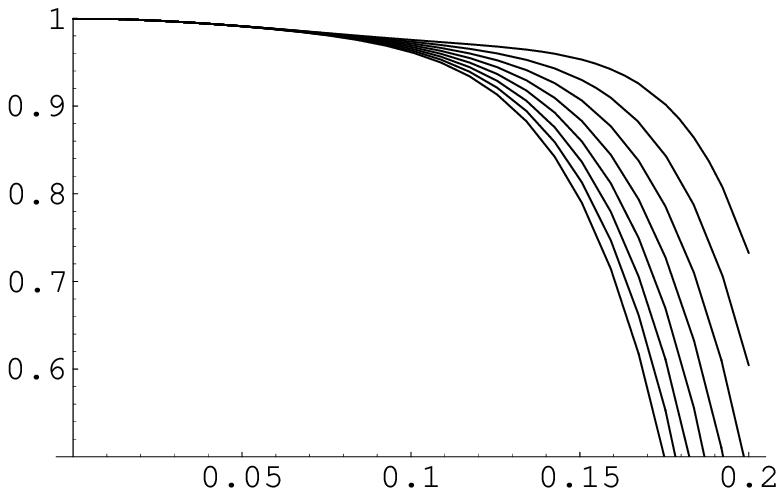}%
\hspace{5mm}%
\epsfxsize=7.4cm\epsfbox{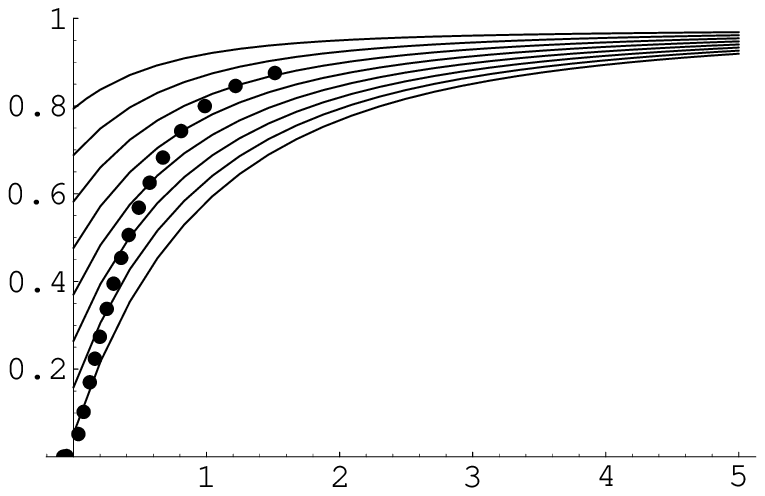}}%
\caption{\label{fig:press}%
{\em Left panel:}
The normalized QCD pressure $p_{\rm QCD}/\pSB$ at $\Nf=0$ plotted versus
the effective coupling $\hhg$ from 
Eq.~(3.3).
The $\hhg^6$ coefficient depends on an unknown parameter $\Delta$
as defined in 
Eq.~(3.8),
and the different curves correspond to choosing $\Delta=-4000$ (lowest curve)
to $\Delta=+10000$, in steps of 2000.
{\em Left panel:}
The same, plotted versus $\ln\frac{T}{T_c}$. The black dots correspond to
lattice data from \cite{Boyd:1996bx}.
} 
\end{figure}

Collecting explicit logarithms $L$, 
they precisely cancel the scale dependence of $\hg^2$ up to 
the order of the computation, and can hence be absorbed by writing
\ba \label{eq:hhg}
\hhg^2 &=& \hg^2 +\hg^4 2\hb_0 L +\hg^6(4\hb_0^2 L^2+2\hb_1 L) 
+\order(\hg^8) \;. 
\ea
Note that this coupling is explicitly scale independent to the 
order we are working, $\partial_{\ln\bar\mu^2}\hhg^2=\order(\hhg^8)$.
We now have the full pressure as a sum of its
ultra-soft, soft and hard parts as 
\ba
p_{\rm QCD} &=& 
d_A \pi^2 T^4 \lb 16 \hat{p}_{\rm us}+16 \hat{p}_{\rm s}
+\frac1{45}\hat{p}_{\rm h} \rb \;,\\
\hat{p}_{\rm us} &=&   
 \hg_M^6 \lk 8\a_G\(\ln\frac{\Nc^2}{4\pi\hg_M^2}+1\) 
 +\frac13 +\pert -\nspt +\nonpert \rk \;,\\ 
\hat{p}_{\rm s} &=& 
 \hm_E^3 \fr13 
 +\hg_E^2\hm_E^2\lk \ln(2\hm_E)-\fr34 \rk
 +\hg_E^4\hm_E\lk -\fr{89}{24}-\fr{\pi^2}6+\fr{11}6\ln2 \rk 
\nn&+&{}
 \hg_E^6\lk \beta_M -8\a_M\ln(2\hm_E) \rk 
 +\hl_E^{(1)}\hm_E^2 \fr{\hn-2}4  
 +\hl_E^{(2)}\hm_E^2 \fr{1-3\hn}4  
\;,\\
\hat{p}_{\rm h} &=&  
\ta_{\rm E1} 
+\hhg^2\ta_{\rm E2} 
+\hhg^4\lk\ta_{\rm E3}-180 \ta_{\rm E5}\rk 
+\hhg^6\lk\tb_{\rm E1} -180\( \tb_{\rm E2} +\ta_{\rm E4}\tb_{\rm E3} 
    +\ta_{\rm E5}\ta_{\rm E7}\) \rk  \;.
\ea
The $g^6$ coefficient of $p_{\rm QCD}$ hence depends on a constant
\ba \label{eq:Delta}
\Delta\equiv\tb_{\rm E1}\pm 720\delta_{\rm NP}\pm 385\delta_{\rm NSPT}\;,
\ea
where we recall that $\tb_{\rm E1}$ stands for the result of the open 4-loop 
computation, and the $\delta$ parameterize the error-bars of the 
numerical constants from \eqss{eq:nonpertRes}{eq:nsptRes} as
$[10.7\pm\delta_{\rm NP}]$ and $[25.8\pm\delta_{\rm NSPT}]$, respectively.
From above, we have $\delta_{\rm NP}=0.4$ and $\delta_{\rm NSPT}=0.8$,
so $\Delta=\tb_{\rm E1}\pm 600$.
In the following, we will for simplicity set 
$\delta_{\rm NP}=\delta_{\rm NSPT}=0$, remembering the induced error-bar 
on $\tb_{\rm E1}$.
Using the same coupling as in $p_{\rm h}$, the above matching conditions 
now read
\ba
\label{eq:mEnew}
 \hm_E^2 &=& 
 \hhg^2 \ta_{\rm E4} 
 +\hhg^4 \ta_{\rm E6}
 +\order(\hhg^6) \,, \\
\label{eq:gEnew}
\hg_E^2 &=& 
 \hhg^2 
 +\hhg^4 \ta_{\rm E7} 
 +\hhg^6 \tilde\gamma_{\rm E1}  
 +\order(\hhg^8) \,,\\
\hl_E^{(1)} &=& 
 \hhg^4 4  +\order(\hhg^6) \;, \\
\hl_E^{(2)} &=& 
 \hhg^4 \fr43(1-z)  +\order(\hhg^6) \;.
\ea

We would now like to plot the result for $p_{\rm QCD}$. 
Identifying the non-interacting (ideal-gas; Stefan-Boltzmann) limit 
as $\pSB=d_A T^4 \frac{\pi^2}{45} \ta_{\rm E1}$, we could display
the normalized pressure $p_{\rm QCD}/\pSB$ as a function of the coupling
$\hhg$, for fixed $\Nf$. 
This is done in the left panel of \fig{fig:press}, at $\Nf=0$ and
for various $\Delta$.
Our goal, however, should be to try to make contact to existing lattice 
determinations of the full pressure, where typically $p_{\rm QCD}/\pSB$
is given as a function of $T/T_c$. Continuum-extrapolated 
lattice data exist for $\Nf=0$ only, 
so in the following we will restrict to this special case.
Aiming for this rather phenomenological comparison, we evidently need
to make some choices, specified below.

\begin{figure}
%
%
%
\centerline{%
\epsfxsize=7.2cm\epsfbox{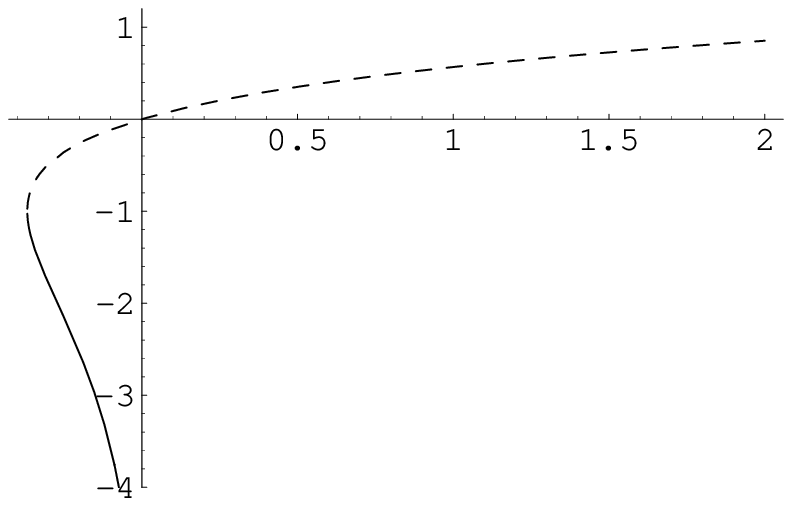}%
\hspace{5mm}%
\epsfxsize=7.2cm\epsfbox{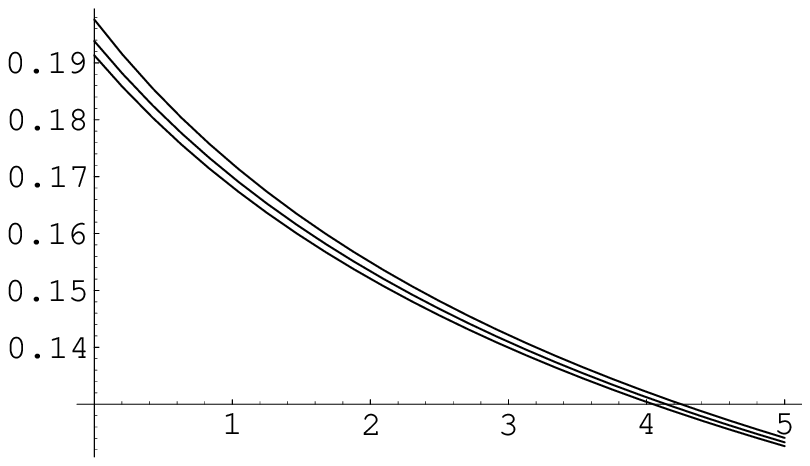}}
\caption{
\label{fig:Wfunction} 
{\em Left panel:} 
The two real branches of $W(z)$ versus $z$. 
The upper (dashed) branch is $W_0(z)$, 
the lower (solid) branch is $W_{-1}(z)$.
{\em Right panel:} 
The effective coupling $\hhg$ from 
Eq.~(3.3)
plotted versus
$\ln\frac{T}{T_c}$, using the choices explained in 
Sec.~3.
The upper/lower curve corresponds to the uncertainty in scale choice 
stemming from fixing $\bar\mu$ and determining $T_c/\lms$ with
$(\delta_{\mu},\delta_{T_c})=(2.0,0.9)/(0.8,1.1)$, 
the bigger effect coming from the latter parameter.} 
\end{figure}

We use the running 4d coupling from the exact solution of 
the 2-loop RGE equation,
\ba \label{eq:alpha2}
\hat g^2(\bar\mu) &=& \frac{-\hb_0/\hb_1}
{1+W_{-1}\(-\frac{\hb_0^2}{\hb_1}
\exp\lk-1-2\frac{\hb_0^2}{\hb_1}\,(L+\ln\frac{4\pi T}{\lms})\rk\)} \;.
\ea
Here, $W_{-1}(z)$ is one of the two real branches of the Lambert W 
function (see e.g. \cite{Wfunction} and the left panel of 
\fig{fig:Wfunction};
$W(z)$ is the function that satisfies $W\exp(W)=z$).
Note that the above solution entails two choices:
The branch of the $W$\/-function and the integration constant were chosen in 
accord with asymptotic freedom (note that the argument of W 
$\rightarrow 0^-$ for $\bar\mu\rightarrow\infty$)
and the `usual' definition of $\lms$ (being the absence of a 
$1/\ln^2\bar\mu$ term in the asymptotic expansion of $\hg(\bar\mu)$ 
at large $\bar\mu$).
Indeed, at large $\hat L=\ln\frac{\bar\mu}{\lms}$ the expansion
$W_{-1}(-\e)=\ln\e-\ln\ln\frac1\e+\order(1/\ln\e)$
reproduces
$
\hat g^2(\bar\mu) = 1/(2\hb_0\hat L+\frac{\hb_1}{\hb_0}\ln(2\hat L)
+\order(\hat L^{-1})) =
\frac1{2\hb_0\hat L}-\frac{\hb_1\ln(2\hat L)}{4\hb_0^3\hat L^2} 
+\order(\hat L^{-3}) \;,
$
in accord with e.g. \re{Eidelman:2004wy}.

Although in principle all dependence on the renormalization scale
$\bar\mu$, entering through $L$, is of higher order, in practice
we need to fix it once we need numerical values for the coupling $\hhg$.
Following \cite{Laine:2005ai}, we
choose the scale $\bar\mu$ by the principle of minimal sensitivity 
applied to the 1-loop result for $\hg_E^2$,
and then estimate the scale-dependence by a variation of a 
factor of $\delta_{\mu}=[0.8..2.0]$ around this $\bar\mu_{\rm opt}$, 
obtaining 
$L=-\frac{\ta_{\rm E7}}{2\hb_0}+\ln\frac{\bar\mu}{\bar\mu_{\rm opt}}$.
The slightly asymmetric choice of $\delta_{\mu}$ here reflects the 
fact that the 1-loop $\hg_E^2$ falls off more steeply on one side of 
the plateau than on the other.

To compare with continuum-extrapolated lattice data \cite{Boyd:1996bx}, 
we use $\frac{T_c}{\lmsbar}=1.22\,\delta_{T_c}$ 
where $\delta_{T_c}=[0.9..1.1]$
encompasses the central values and error bars of estimates of this quantity 
from different lattice collaborations (for a summary of the different 
methods and results, see \cite{Laine:2005ai}). 
This would translate into a horizontal error bar for the lattice data
when plotted against $T/\lmsbar$.


In the right panel of \fig{fig:Wfunction}, we have plotted 
the effective coupling 
$\hhg$ as defined in \eq{eq:hhg}, converted to a function of 
$\ln\frac{T}{T_c}$. Note that its value is smaller than $0.2$ even
at $T_c$.

The normalized pressure $p_{\rm QCD}/\pSB$, converted to a function of
$\ln\frac{T}{T_c}$ along the lines above, 
is displayed in the right panel of \fig{fig:press}.
For comparison, the continuum-extrapolated lattice data of 
\re{Boyd:1996bx} has been included as black dots. 
The figure suggests a value for the $\Nf=0$ coefficient of the
unknown constant $\tb_{\rm E1}$, $\#_0\approx 6000\pm 600$, bearing in mind
the error bar defined in \eq{eq:Delta}.


\section{Discussion}

We have currently no idea what the 4-loop hard-scale coefficient
$\tb_{\rm E1}$ is, even though it can be computed diagrammatically.
As already mentioned above, it should be a polynomial
in $z=\Nf/\Nc$, $\tb_{\rm E1}=\#_0+z\#_1+z^2\#_2+z^3\#_3$,
where only a single (ring) diagram contributes to $\#_3$, suggesting
it as the first test-case for 4-loop sum-integral technology.

It seems possible to give an estimate of the highest-$\Nf$ contribution 
to $\tb_{\rm E1}$ from the large-$\Nf$ solution for the pressure,
since terms of order $g^6 \Nf^3$ originate from the hard-scale pressure 
$p_{\rm E}$ only.
In \cite{Ipp:2003jy} this was attempted by fitting 
the numerically known exact large-$\Nf$ pressure with a polynomial in $g$.
This results in 
$\#_3=\frac{45}{8\pi^2}[+20(2)]-\frac{20}9\ln2\(1+12\gamma_0-\frac{364}5\ln2+16Z_1-8Z_3\)\approx[+36(1)]$,
where the terms proportional to $\ln2$ originate from translating 
the choice of renormalization scale $\bar\mu=\pi T$ of \cite{Ipp:2003jy}
to our definition of $\tb_{\rm E1}$, where powers of 
$L=\ln\frac{\bar\mu}{4\pi T}$ were subtracted out.

Furthermore, fitting the full $g^6$ pressure at $\Nf=0$ and $\Nc=3$ 
to lattice data around $4T_c$ \cite{Boyd:1996bx} suggests
a value $\#_0\approx6000\pm 600$,
if one takes the conjecture for granted that all higher-order corrections
sum up to a subdominant contribution.
There is no guarantee whatsoever that this conjecture holds,
making a perturbative computation of the $\#_i$ unavoidable.
We take the above check against the lattice data as indication
that the effective theory setup has a chance to analytically
describe the transition from temperatures as low as a few times $T_c$
to infinite temperatures, in terms of computable corrections to
the ideal-gas limit.


\acknowledgments

Y.S. would like to thank the organizers for a very enjoyable workshop,
A.~Patk\'os for sharing his insights into Hungarian art,
and M.~Laine for valuable comments on the manuscript.


\end{document}